\documentstyle[pre,aps,epsf,psfig]{revtex}
\newcommand \be[1]{\begin{equation}\label{#1}}
\newcommand \ee{\end{equation}}
\newcommand \beq[1]{\begin{eqnarray}\label{#1}}
\newcommand \eeq{\end{eqnarray}}
\newcommand \bib{\bibitem}

\newcommand{\DK}[1]{\mbox{\boldmath$#1$}}

\newcommand \om{\omega}
\newcommand \eps{\epsilon}

\begin{document}

\draft
\title{
Relativistic photoelectron spectra in the ionization of atoms by elliptically polarized light
}

\author{J. Ortner}
\address{
{\it Institut f\"ur Physik,Humboldt Universit\"at zu Berlin, 
Invalidenstr. 110, 10115 Berlin, Germany}
}
\date{\today}
\maketitle
\begin{abstract}
Relativistic tunnel ionization of atoms by intense, elliptically polarized light is considered. The relativistic version of the Landau-Dykhne formula is employed. The general analytical expression is obtained for the relativistic photoelectron spectra. The most probable angle of electron emission, the angular distribution near this angle, the position of the maximum and the width of the energy spectrum are calculated. In the weak field limit we obtain the familiar  non-relativistic results. For the case of circular polarization our analytical results are in agreement with recent derivations of Krainov [V.P. Krainov, J. Phys. B, {\bf 32}, 1607 (1999)]. 
\end{abstract}


\section{Introduction}

Recently an increasing interest in the investigation of relativistic ionization phenomena has been observed \cite{R90,CR94,PMK97,PMK98,DK98,K98,CR98,CR99,K99,OR99}. Relativistic effects will appear if the electron velocity in the initial bound state or in the final state is comparable with the speed of light. The initial state should be considered relativistic in the case of inner shells of heavy atoms \cite{PMK97,PMK98}. In a recent paper \cite{OR99} the photoionization of an atom from a shell with relativistic velocities has been considered for the case of 
elliptically
polarized laser light. In the present paper we will study the effect of a relativistic final-state of an electron on the ionization of an atom by elliptically polarized light. The initial state will be considered as nonrelativistic. The final-state electron will have an energy in the laser field measured by the ponderomotive energy. If the ponderomotive energy approaches the electron rest energy, then a relativistic treatment of the ionization process is required. For an infrared laser the necessary intensities are of the order of $10^{16} {\rm W}\,{\rm cm}^{-2}$.

Ionization phenomena influenced by relativistic final state effects have been studied for the cases of linearly and circularly polarized laser radiation both in the tunnel \cite{K98,K99} and above threshold regimes \cite{CR94,CR98}. The ionization rate for relativistic electrons has been found to be very small for the case of linear polarization \cite{K98,CR98}. On the contrary a circularly polarized intense laser field produces mainly relativistic electrons \cite{CR94,K99}.

In the papers of Reiss and of Crawford and
Reiss  \cite{R90,CR94,CR98} a covariant version of the so-called strong field approximation \cite{R80} has been given for the cases of linear and circular polarization.
 Within this approximation one calculates the transition amplitude between the initial state taken as the solution for the Dirac equation for the hydrogen atom and the final state described by the relativistic Volkov solution. Coulomb corrections are neglected in the final Volkov state.  Analytical
results for the ionization rate have been given in Refs. \cite{R90,CR94,CR98} These results apply  to above barrier cases as well as to tunneling cases.  However, the corresponding expressions are complicated and numerical calculations are needed to present the final results. 

The present paper is aimed to 
investigate the relativistic electron energy spectra in the ionization of atoms by intense {\em elliptically} polarized laser light. 
In contrast to the more sophisticated strong field approximation we would like to obtain simple analytical expressions from which the dependence of the ionization process on the parameters, such as binding energy of the atom, field strength, frequency and ellipticity of the laser radiation may be understood without the need of numerical calculations. Therefore we restrict the considerations to the case of tunnel ionization. Our results will be applicable only for laser field strengths smaller than the inner atomic field $F \ll F_a$. In order to observe relativistic effects, the inequality $\eps=F/\om c > 0.1$ should be fulfilled. (The atomic system of units is used throughout the paper, $m=e=\hbar=1$.) Both inequalities yield a limitation for the laser frequency $\om$ from above. For the ionization of multi-charged ions an infrared laser satisfies this condition. 

The non-relativistic sub-barrier ionization with elliptically polarized light was studied in \cite{PPT66}. In the tunnel limit the simple expression
\be{nonr}
W^{nonrel} \propto \exp \left\{ - \frac{4}{3}\frac{\gamma}{\om}E_b \left[1-\frac{1}{10}\left(1-\frac{g^2}{3}\right)\right] \right\} \exp \left\{ -\frac{\gamma}{\om} \left[ \left(p_z-g \frac{F}{\om} \right)^2  + p_x^2 \right] \right\}
\ee
has been derived for the electron momentum spectrum within exponential accuracy. Here $p_x$ and $p_z$ are the projections of the drift momentum on the direction of the wave propagation and along the smaller axis of the polarization ellipse, respectively; $E_b$ is the ionization energy of the atomic state, $F$, $\om$ and $g$ are the field amplitude, frequency and ellipticity of the laser radiation, respectively;  and $\gamma = \om \sqrt{2 E_b}/F \ll 1$ is the Keldysh adiabatic parameter.

From Eq. (\ref{nonr}) one concludes that the electrons are mainly ejected in the polarization plane along the smaller axis of polarization; the most probable momentum at the time of ejection has the components: $p_x=p_y=0$ and $p_z=gF/\om$. (For the sake of simplicity of the notations we neglect througout the paper the second symmetric maximum for the component $p_z$, $p_z=-gF/\om$.)

\section{Relativistic semiclassical approach}

We shall now generalize the non-relativistic result Eq. (\ref{nonr}) to the case of relativistic final state effects, when $g F/\om$ becomes comparable with the velocity of light. Our derivation starts with the relativistic version of the Landau-Dykhne formula \cite{PMK97,DK98,OR99}. The ionization probability in quasiclassical approximation and with exponential accuracy reads
\be{prob}
{W} \propto \exp\left\{- {2} \, {\rm Im}~\left(S_f(0;t_0)+S_i(t_0)\right)\right\} \,,
\ee
where $S_i=-E_0 t_0$ is the initial part of the action, $S_f$ is the final-state action. In the latter we will neglect the influence of the atomic core. Then the final-state action  may be found as a solution of the Hamilton-Jacobi equation and reads \cite{Landau}
\be{Srel}
S_f(0;\xi_0)=c \Biggl\{ r \xi_0 + \frac{\eps c}{q \om} \left[ -p_y \cos \om \xi_0 - p_z g \sin \om \xi_0 \right] +\frac{\eps^2 c^2}{4 q} \left[ \left(1+g^2) \xi_0 + \frac{g^2-1}{2 \om} \sin 2 \om \xi_0 \right] \right\}\,.
\ee
Here the vector potential of the laser radiation has been chosen in the form 
\be{pot}
A_x=0\,,~~~A_y=-\frac{cF}{\om} \sin \om \xi\,,~~~A_z=g \frac{cF}{\om} \cos \om \xi\\,,
\ee
where $\xi=t-{x}/{c}$, $\xi_0$ is the initial value. Further the notations
\be{not}
r = \sqrt{c^2+p^2}\,,~~~~~~q=r-p_x\,
\ee
have been introduced; $p_x$, $p_y$ and $p_z$ are the components of the final electron momentum along the beam propagation, along the major and along the small axis of the polarization ellipse, respectively; $p^2=p_x^2+p_y^2+p_z^2$.

 The complex initial time $t_0$ has to be determined from the classical turning point in the complex half-plane \cite{PMK97,DK98,K99,OR99}:
\be{cond1}
E_f(t_0)=c\Biggl\{ r  + \frac{\eps c}{q} \left[p_y \sin \om t_0 - g p_z \cos \om t_0 \right] +   \frac{\eps^2 c^2}{2 q} \left[ \frac{g^2+1}{2} + \frac{g^2-1}{2} \cos 2 \om t_0 \right]  \Biggr\}=E_0= c^2-E_b \,.
\ee
Eq. (\ref{prob}) together with Eqs. (\ref{Srel}) and (\ref{cond1}) is the most general expression for the relativistic rate of sub-barrier ionization by elliptically polarized laser light. We consider now the limit of a nonrelativistic initial state, i.e. $E_b \ll c^2$. Furthermore the considerations will be restricted to the tunnel regime $\lambda=-i \om t_0 \ll 1$, or, equivalently, the Keldysh adiabatic parameter should satisfy the inequality $\gamma  \ll 1$. Under these conditions we may expand the sine and cosine functions in Eqs. (\ref{Srel}) and (\ref{cond1}) in Taylor series. 
Then we obtain the rate of tunnel ionization for arbitrary final-state momenta.
Expanding this expression near its maximum value in terms of the parameters $q$, $p_y$ and $p_z$ one arrives at the following general expression
\be{rel2}
W^{rel} \propto \exp \left\{- \frac{4}{3}\frac{\gamma}{\om}E_b \left[1-\frac{\gamma^2}{10}\left(1-\frac{g^2}{3}\right)-\frac{E_b}{12 c^2} \right] \right\} \exp \left\{ -\frac{\gamma}{\om} \left[ \left(p_z-p_{z,m} \right)^2  + \left(q-q_m \right)^2 \right] \right\}
\ee
for the tunnel ionization rate (first exponent) and the momentum distribution of the photoelectron (second exponent) within exponential accuracy. In Eq. (\ref{rel2}) the ionization rate and the most probable value for each component of the electron momentum,
\be{momenta}
p_{y,m}=0\,~~~p_{z,m}= \frac{F}{\om} g \left(1+\frac{\gamma^2}{6}\right)\,,~~~~~~q_m=c-\frac{E_b}{3 c}
\ee
are given including the first frequency and relativistic corrections in the initial state. In the distribution near the maximum momenta only those terms have been maintained which do not vanish at zero frequency.  Equation (\ref{rel2})  agrees with the relativistic angular-energy distribution of Krainov \cite{K99} in the case of circular polarization $g = \pm 1$, vanishing frequency corrections $\gamma^2 \ll 1$ and negligible relativistic effects in the initial state $E_b \ll c^2$. In the nonrelativistic limit,i.e., $p \ll c$ , $F/w c \ll 1$ and $E_b \ll c^2$, we have $q-q_m=p_x$ and Eq. (\ref{rel2}) reduces to Eq. (\ref{nonr}) as it should. 

From Eq. (\ref{momenta}) we easily obtain the most probable value for the component of the electron momentum along the beam propagation
\be{p_x}
p_{x,m}=\frac{F^2 g^2}{2 \om^2 c}+\frac{E_b}{3 c}\left(g^2+1\right)\,,
\ee
the peak value of the angular distribution
\be{angle}
\tan \theta_m=\frac{p_{x,m}}{|p_{z,m}|}=\frac{F|g|}{2c\om} \left(1+\frac{g^2+2}{g^2}\frac{\gamma^2}{6} \right)\,,~~~\varphi_m=0\,,
\ee
and the value of the most probable electron energy $E_m=p_m^2$, with
\be{energy}
p_m=\sqrt{p_{x,m}^2+p_{z,m}^2}=\frac{F|g|}{\om}\sqrt{1+\left(\frac{Fg}{2\om c}\right)^2+\frac{\gamma^2}{3}+\frac{E_b}{3c^2}(g^2+1)}\,.
\ee
Here $\theta$ is the angle between the polarization plane and the direction of the photoelectron motion; $\varphi$ is the angle between the projection of the electron momentum onto the polarization plane and the smaller axis of the polarization ellipse. For the ellipticity $0<|g|<1$ the most probable momentum $\DK{p_m}$ of the ejected electron is situated in the plane perpendicular to the maximum value of the electric field strength; for $|g|=1$ the electron output in the $(y,z)$ plane is isotropic. Notice that the most probable total electron momentum $p_m$ contains relativistic final state effects, frequency corrections and weak relativistic initial state effects. Relativistic effects do not contribute to the projection of the momentum along the smaller axis of the polarization ellipse. On the contrary both relativistic final and initial state effects  increase the electron momentum projection along the propagation of elliptically polarized laser radiation. The increase due to relativistic initial state effects is proportional to $(E_b/c^2) (1+g^2)$. It is typically small (except for the  ionization from K shells of heavy atoms \cite{PMK97,OR99}) and does not vanish in the case of linear polarization of the laser light. In contrast to that the relativistic increase due to final state effects which is measured by ${Fg}/{2\om c}$ is absent in the case of linear polarized laser radiation.

In what follows  we will neglect  the frequency corrections and the relativistic initial state effects in order to compare with previous works. In this case and for the case of circular polarization the expressions for the angle $\theta_m$ and the most probable electron momentum $p_m$ coincide with the corresponding expressions of Krainov \cite{K99}. Moreover, though our calculations are valid only in the tunnel regime, our value for the most probable  angle of electron ejection coincides with an approximation given by Reiss for the case of circular polarization \cite{R90,CR94} and valid in the above-barrier ionization regime. In Ref. \cite{CR94} it has been shown that the simple estimate $\tan \theta_m = {F}/{2c\om}$ is in good agreement with the  numerical calculations based on the strong-field approximation  and performed for above threshold conditions with circularly polarized light. Therefore we expect that our formula Eq. (\ref{angle}) predicts,  at least qualitatively, the location of the peak in the relativistic angular distribution for the case of above barrier ionization with elliptically polarized light.  

This statement is supported by a semiclassical consideration of the above barrier ionization. According to the semiclassical model \cite{CBB92} the transition occurs from the bound state to that continuum state which has zero velocity at the time $t$ with the phase $\xi$ of the vector potential $\DK{A}(\xi)$.  From this condition we have
\beq{semi}
q&=&\sqrt{c^2+p_y^2+p_z^2+ \eps^2 c^2 g^2 + 2 \eps c \left(p_y \sin \om \xi -g p_z \cos \om \xi \right) + (1-g^2) \eps^2 c^2 \sin^2 \om \xi} \,,\\  
p_y&=&-\frac{F}{\om} \sin \om \xi\,,~~~
p_z=g \frac{F}{\om} \cos \om \xi\,.
\eeq
The ionization rate becomes maximal at the maximum of the electric field of the laser beam. 
Due to our choice of the gauge (see Eqs. (\ref{pot})) this maximum occurs at the phase $\xi=0$. From Eqs. (\ref{semi}) and the relation $p_x=(c^2-q^2+p_y^2+p_z^2)/2q$ we conclude that the most probable final state has the momentum with the components
\be{semip}
p_x= \frac{F^2 g^2}{2 c \om^2}\,,~~~p_y=0\,,~~~p_z=\frac{F}{\om}\,,
\ee
which agrees with the above estimations Eqs. (\ref{momenta}) and (\ref{p_x}) derived for the case of tunnel ionization if one neglects the frequency corrections and the relativistic initial state effects.

\section{Results and conclusions}

It is now straightforward to obtain the probability distribution for the components of the final state momentum. Neglecting again the frequency corrections and the relativistic initial state effects we get from Eq. (\ref{rel2}) 
\be{comp}
W^{rel} \propto \exp \left\{-\frac{4}{3}\frac{\gamma}{\om}E_b  \right\} \exp \left\{ -\frac{\gamma}{\om} \frac{\left[ \delta p_x ^2  -2 \delta p_x \delta p_z \eps g + p_y^4/4c^2 + \delta p_z^2 \left(1+2\eps^2 g^2 +{\eps^4 g^4}/{4}\right)\right]}{\left(1+\eps^2 g^2/2 \right)^2} \right\}\,,
\ee
where only the leading contributions in $\delta p_x=\left(p_x-p_{x,m} \right)$, $p_y$ and $\delta p_z=\left(p_z-p_{z,m} \right)$ have been given. In the non-relativistic limit $\eps \ll 1$ and $p \ll c$ we obtain Eq. (\ref{nonr}). For the case of linear polarization $g=0$ we reproduce the momentum distribution of Krainov \cite{K98} including the relativistic high energy tail for electrons emitted along the polarization axis. The latter is described by the term $\exp\left\{-(\gamma/\om) (p_y^4/4c^2)\right\}$. However, the high energy tail contains only a very small part of the ejected electrons. From Eq. (\ref{comp}) we see that in the case of linear polarization most of the electrons have nonrelativistic velocities. This is in agreement with recent numerical calculations based on the strong-field approximation \cite{CR99}. In contrast to the case of linear polarized laser radiation, the intense elliptically polarized laser light with $\eps |g|$ of the order of unity produces mainly relativistic electrons.

For the sake of comparison we shall give the angular distribution at the maximum of the electron energy spectrum and the energy spectrum at the peak of the angular distribution. We obtain both distributions from Eq. (\ref{rel2}) by putting $p_x=p \sin \theta$ and $p_z=p \cos \theta$, where we have taken into account that the ionization rate is maximal for the emission in the $(x,z)$ plane. Choosing the peak value of the angular distribution $\theta=\theta_m= {\rm{arctan}}\; \eps |g|/2$ we obtain
\be{energy-an}
W^{rel} \propto \exp \left\{ - \frac{2}{3} \frac{\left(2 E_b\right)^{3/2}}{F}\right\} \exp \left\{ -\left(\frac{p-p_m}{\Delta p}\right)^2 \right\}
\ee
for the energy distribution along the most probable direction of electron ejection. Here 
\be{width1}
\Delta p =\sqrt{\frac{F}{\sqrt{2 E_b}}}\frac{1+({g^2}/{2})\left({F}/{\om c}\right)^2}{\sqrt{1+g^2\left({F}/{\om c}\right)^2}}\,,
\ee
is the width of the relativistic energy distribution. From Eq. (\ref{width1}) we conclude that the relativistic width is broader than the nonrelativistic one, it increases with increasing field strength. The relativistic broadening has its maximum for circular polarization, there is no relativistic broadening of the energy width for the case of linear polarization. 

In Fig. 1 the electron momentum spectrum from Eq. (\ref{energy-an}) is shown  for electrons born in the  creation of  ${\rm Ne}^{8+}$ ($E_b=239\, {\rm eV}$) ions by an elliptically polarized laser radiation with wave length $\lambda=1.054 \,\mu {\rm m}$, field strength $2.5 \times 10^{10} \,{\rm V/cm}$ and ellipticity $g=0.707$. The relativistic spectrum is  compared with the spectrum of nonrelativistic theory. From the figure one sees the shift of the energy spectrum to higher energies, the relativistic broadening of the spectrum is too small to be observed from the figure.

Putting in equation (\ref{rel2}) $p=p_m=({F|g|}/{\om})\sqrt{1+\left({Fg}/{2\om c}\right)^2}$ we obtain the angular distribution for the most probable photoelectron energy,
\be{angular}
W^{rel} \propto \exp \left\{ - \frac{2}{3} \frac{\left(2 E_b\right)^{3/2}}{F}\right\} \exp \left\{ -  \left(\frac{\theta-\theta_m}{\Delta \theta}\right)^2 \right\}\,,
\ee
where the width of the angular distribution equals
\be{width2}
\Delta \theta= \frac{\om}{|g|} \sqrt{\frac{1}{F\sqrt{2 E_b}}} \frac{1}{\sqrt{1+g^2 (F/2 \om c)^2}}\,.
\ee
We see that the relativistic theory predicts a narrower angular distribution as the nonrelativistic theory. The infinite width in the case of linear polarization is an artefact of the calculations using the angle between the polarization plane and the direction of electron movement. For the linear polarization the electrons are ejected preferably along the polarization axis if one neglects relativistic initial state effects. For the case of circular polarization the energy-angular distributions Eqs. (\ref{energy-an}) and (\ref{angular}) coincide with the corresponding expressions of Krainov \cite{K99}. Notice that our notations slightly differ from those of Krainov.

In Fig. 2 we have plotted the  relativistic and non-relativistic angular distributions for the electrons produced by the same process as in Fig. 1. The relativistic distribution has its maximum at the angle $\theta_m={\rm{arctan}}\; \eps |g|/2=16.71^{\circ}$ and the nonrelativistic theory predicts a peak at the zero angle. Again the relativistic reduction of the angular distribution width is not observable for the parameters we have chosen. From Figs. 1 and 2 one concludes that the appearance of a nonzero mean  of the drift momentum component along the beam propagation and the shift of the mean emission angle into the forward direction are the most important indications for a relativistic ionization process. On the contrary the width of the energy-angle distributions as well as the total ionization rate are less sensitive to the relativistic final state effects.

In conclusions, in this paper the relativistic semiclassical ionization of an atom in the presence of intense elliptically polarized laser light has been considered. Simple analytic expressions for the  relativistic photoelectron spectrum have been obtained.  For the cases of linear and circular polarization our results agree with previous studies. We have shown that the location of the peak in the relativistic angular distribution is shifted toward the direction of beam propagation. The theoretical approach employed in the paper predicts that the maximum of the electron energy spectrum is increased due to relativistic effects. The validity of the simple expressions is formally limited to the tunnel regime. Nevertheless, a part of the results, such as the most probable angle for electron emission, is shown to be valid in the above barrier ionization regime as well. The results obtained in this paper within exponential accuracy may be improved by the account of Coulomb corrections. However, whereas the Coulomb corrections may strongly influence the total ionization rate \cite{ADK86,BM99} we expect only a small influence of the atomic core on the electron spectrum.

\section{Acknowledgements}
I gratefully acknowledge usefull discussions with V.M. Rylyuk. This work is financially supported by the Deutsche Forschungsgemeinschaft (Germany) under Grant No. Eb/126-1.

\newpage

\begin{center}
{\bf FIGURE CAPTIONS}
\end{center}

\begin{description}

\item[(Figure 1)] Electron momentum spectra for electrons produced in the creation of ${\rm Ne}^{8+}$ by an elliptically polarized laser radiation with wave length $\lambda=1.054 \,\mu {\rm m}$, field strength $2.5 \times 10^{10} \,{\rm V/cm}$ and ellipticity $g=0.707$ and ejected at the most probable angle $\theta=\theta_m $; the relativistic spectrum is taken from Eq. (\ref{energy-an}) with $\theta_m=16.71^{\circ}$, the non-relativistic one from Eq. (\ref{nonr}) with $\theta_m=0$.
\item[(Figure 2)] Electron angular distribution at the most probable electron momentum $p=p_m$. The other parameters are the same as in Fig. 1; the relativistic angular distribution is taken from Eq. (\ref{angular}) with $p_m=85.91$, the non-relativistic one from Eq. (\ref{nonr}) with $p_m=82.28$.

\end{description}

\newpage

\begin {figure} [h] 
\unitlength1mm
  \begin{picture}(155,160)
\put (0,10){\psfig{figure=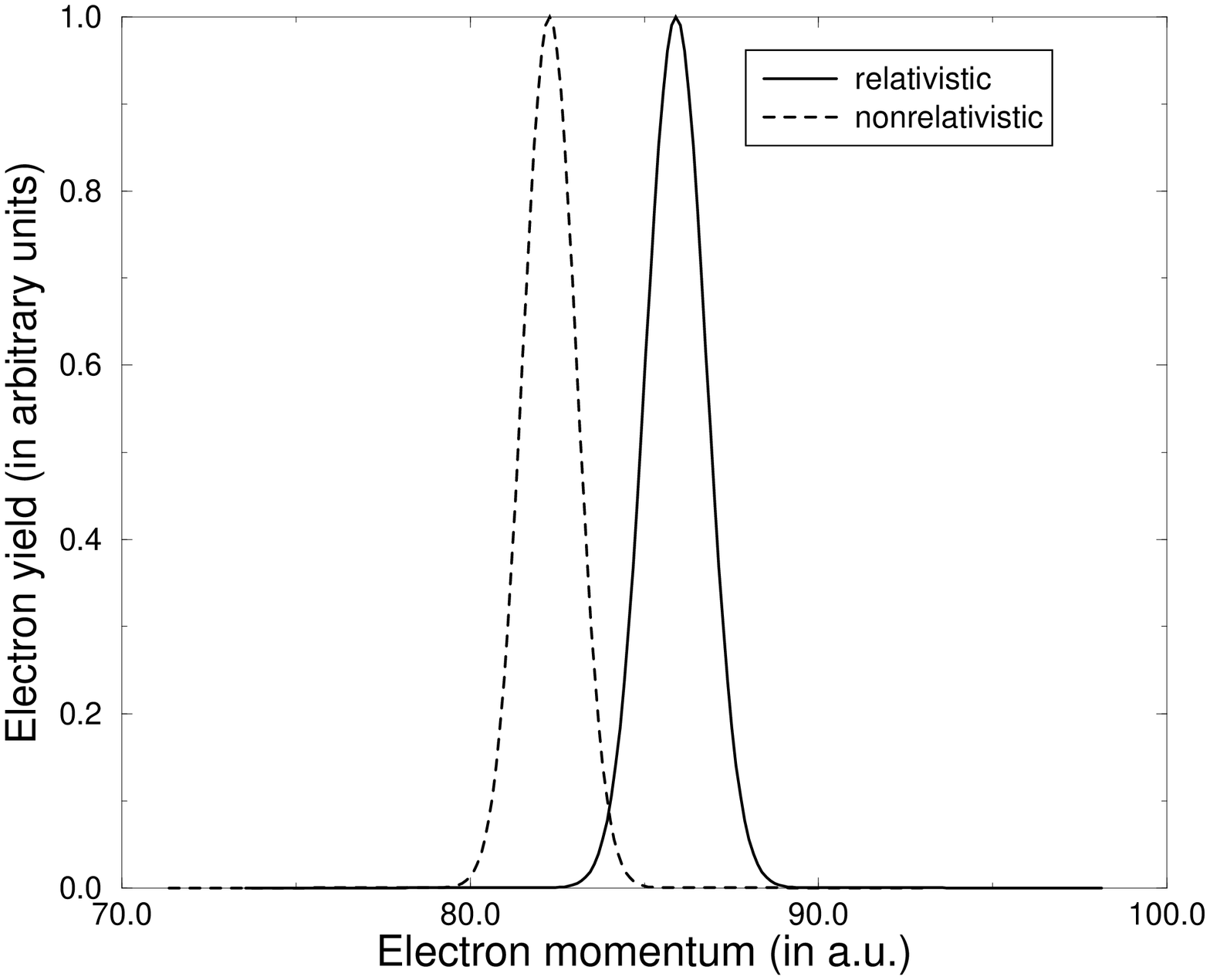,width=15.0cm,height=13.0cm,angle=-90}}
 \end{picture}\par
\caption{}
\end{figure}

\newpage

\begin {figure} [h] 
\unitlength1mm
  \begin{picture}(155,160)
\put (0,10){\psfig{figure=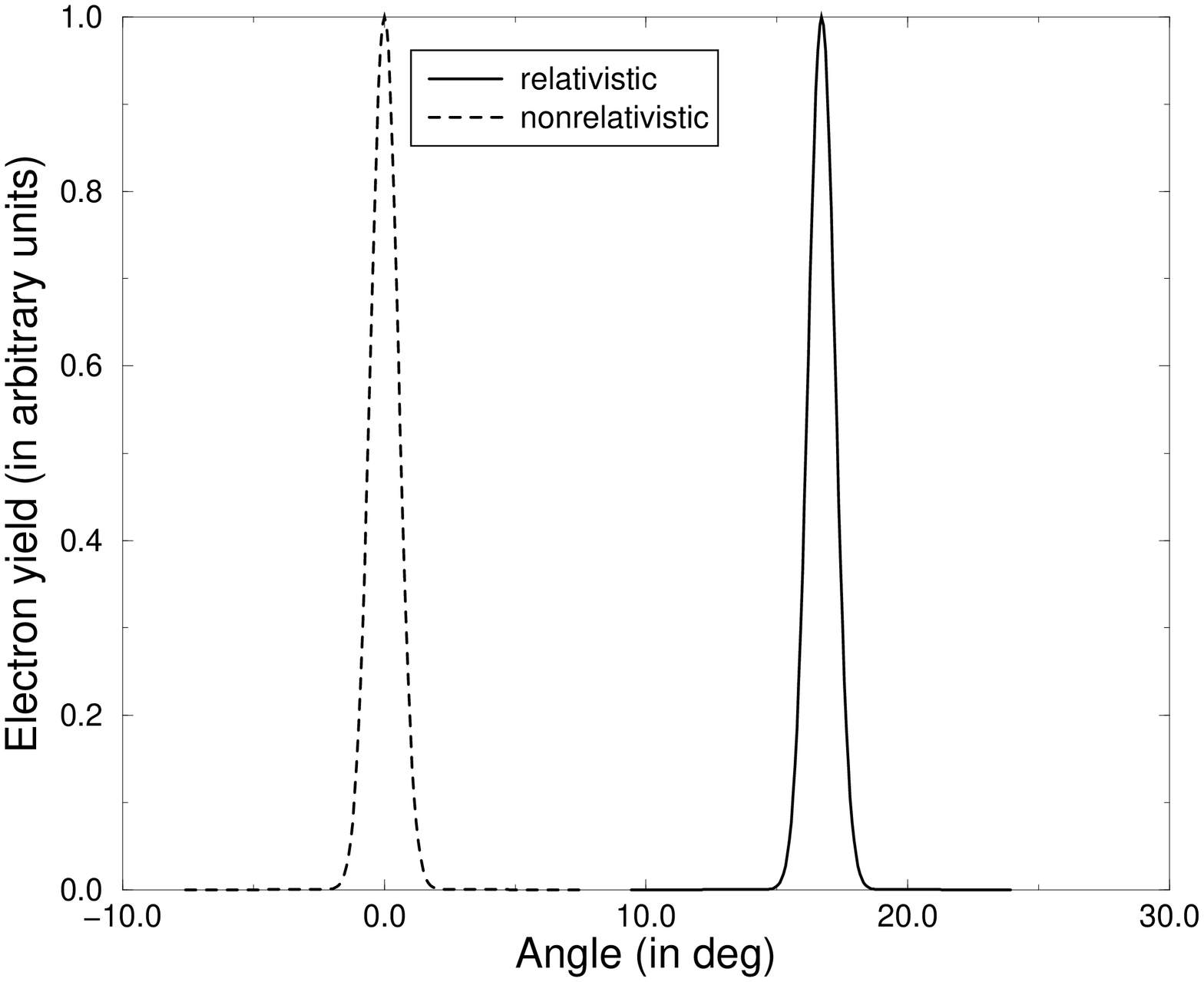,width=15.0cm,height=13.0cm,angle=-90}}
 \end{picture}\par
\caption{}
\end{figure}

\end{document}